\documentclass[conference,a4paper]{APSIPA2021}
\usepackage{multirow}
\usepackage{graphicx}
\usepackage{amsmath}
\usepackage[psamsfonts]{amssymb}
\usepackage{amsxtra}
\usepackage{threeparttable}

\usepackage{amsfonts}
\usepackage{xcolor}
\usepackage{booktabs, makecell}
\usepackage{color}
\usepackage{colortbl}
\definecolor{myblue}{RGB}{240,247,255}
\definecolor{mygray}{RGB}{242,242,242}
\definecolor{mygreen}{RGB}{248,252,246}
\usepackage{caption}
\usepackage{subcaption}
\usepackage{algorithmic}
\usepackage[ruled]{algorithm2e}
\usepackage{url,cite}

\makeatletter
\newcommand*{\rom}[1]{\expandafter\@slowromancap\romannumeral #1@}
\makeatother

\usepackage{geometry}
\usepackage{fancyhdr}

\begin{document}
\title{Disentangled Speaker Representation Learning\\via Mutual Information Minimization}

\author{
\authorblockN{%
Sung Hwan Mun\authorrefmark{1}, Min Hyun Han\authorrefmark{1}, Minchan Kim, Dongjune Lee, and Nam Soo Kim 
}
\authorblockN{ 
Department of Electrical and Computer Engineering and INMC, Seoul National University, Seoul, South Korea \\
\{shmun, mhhan, mckim, djlee\}@hi.snu.ac.kr, nkim@snu.ac.kr} \\
}

\maketitle
\begingroup\renewcommand\thefootnote{\authorrefmark{1}}
\footnotetext{Equal contribution.}

\begingroup\renewcommand\thefootnote{1}

\begin{abstract}
Domain mismatch problem caused by speaker-unrelated feature has been a major topic in speaker recognition. In this paper, we propose an explicit disentanglement framework to unravel speaker-relevant features from speaker-unrelated features via mutual information (MI) minimization. To achieve our goal of minimizing MI between speaker-related and speaker-unrelated features, we adopt a contrastive log-ratio upper bound (CLUB), which exploits the upper bound of MI. Our framework is constructed in a 3-stage structure. First, in the front-end encoder, input speech is encoded into shared initial embedding. Next, in the decoupling block, shared initial embedding is split into separate speaker-related and speaker-unrelated embeddings. Finally, disentanglement is conducted by MI minimization in the last stage. Experiments on Far-Field Speaker Verification Challenge 2022 (FFSVC2022) demonstrate that our proposed framework is effective for disentanglement. Also, to utilize domain-unknown datasets containing numerous speakers, we pre-trained the front-end encoder with VoxCeleb datasets. We then fine-tuned the speaker embedding model in the disentanglement framework with FFSVC 2022 dataset. The experimental results show that fine-tuning with a disentanglement framework on a existing pre-trained model is valid and can further improve performance.
\end{abstract}

\section{Introduction}
Speaker verification is a task of determining whether the input speech is spoken by the same speaker or not \cite{15Hansen}. The general speaker verification framework consists of an embedding extraction and scoring process. In the embedding extraction step, audio with variable duration is converted into a single fixed-dimensional vector representation called speaker embedding, which is assumed to contain speaker-relevant information. With a sophisticated speaker embedding, even a simple scoring method such as cosine similarity or euclidean distance has shown high speaker verification performance \cite{jung2022pushing, brown2022voxsrc, kwon2021ins}. Therefore, most studies have been focused on how to extract a fine speaker embedding from input speech.

With the development of the deep learning field, various studies have been proposed to utilize the neural network for extracting speaker representation called deep speaker embedding, which reflects the speaker's characteristics well\cite{17Li, 18Snyder, 18Okabe}. Despite the success of deep speaker embedding methods, there still remains the problem of performance degradation in mismatched conditions (e.g., device, noise, language). In order to solve this problem, there has been a demand for robust speaker embedding, unaffected by the domain mismatch due to speaker-irrelevant factors.

Traditionally, data augmentation is the most common approach for training neural networks robust to domain mismatch. For speaker verification, simulated reverberation \cite{17Ko}, additive noise \cite{15Snyder}, and SpecAugment \cite{Park20Specaug} can be good options for data augmentation to increase the number of acoustic environments that might be encountered in the inference phase \cite{20Heo, 20Desplanques}. While these methods are proven to be effective when there are insufficient data on target conditions, they can only indirectly mitigate the domain mismatch problem.

Unlike the methods described above, various studies have been proposed to disentangle speaker-irrelevant variability from the speaker embedding directly. Recently, adversarial learning-based domain adaptation methods have been studied. \cite{19ZMeng, 21Wang, Huh20AAT} utilized gradient reversal layer (GRL) to prevent speaker embeddings network from learning the information needed for the sub-task (i.e., noise classification). Although the gradient reversal techniques have proven to be effective for performance improvement, training a network with GRL is known to be unstable and sensitive to the hyper-parameter setting. As an alternative to GRL, domain adversarial training similar to the generative adversarial network (GAN) framework was exploited to maximize the error on the subtask \cite{19Bhattacharya, 19Zhou}. However, these domain adaptation methods have a limitation that adaptation is applied to the feature space shared by both speaker-relevant and speaker-irrelevant factors. Therefore, speaker embedding is inherently hindered by speaker-independent factors. Also, adversarial training has known to be difficult and unstable \cite{Kang20JFE}.

Alternatively, there have been several approaches to minimize correlation between speaker and speaker-independent embeddings in distinct space. For instance, joint factor embedding (JFE) \cite{Kang20JFE} framework simultaneously extracts speaker and nuisance (i.e., non-speaker) embeddings and maximizes entropy (or uncertainty) on their opposite task, while minimizing correlation between two embeddings using mean absolute Pearson's correlation (MAPC) computed batch-wise. Similarly, \cite{19Tai, 20Kwon}  divided features into the speaker and residual embeddings and increased their uncertainty on the contrary task, and \cite{Sang21DEEAN} minimized mutual information via mutual information neural estimator (MINE) with GRL. Additionally, they adopted an autoencoder framework for training merged embedding to maintain the complete information of input speech \cite{19Tai, 20Kwon, Sang21DEEAN}. However, naively increasing uncertainty on the other task does not guarantee disentanglement.

For learning disentangled representations, mutual information (MI) minimization has gained considerable interest in various machine learning tasks \cite{05Babaie-Zadeh, 21Zhu, 21Hou}. Since the exact computation of MI in high-dimensional space is intractable when only sample-based approaches are available, several prominent MI estimators have been proposed \cite{16Chen, Oord18CPC, 18Belghazi, 20Cheng}. Among them, contrastive log-ratio upper bound (CLUB) \cite{20Cheng} proposed the MI upper bound estimator by using the difference of conditional probabilities between positive and negative sample pairs in a contrastive learning manner. As our goal is to learn disentangled speaker embedding, we utilize CLUB to reduce the interdependence between opposite latent representations explicitly.

In this work, we propose an effective learning framework for disentangled speaker representation via MI minimization. To learn speaker embedding that is not only soundly disentangled but also has high speaker discrimination ability, we construct a 3-stage structure; \textit{Front-end Encoder Network}, \textit{Decoupling Block}, and \textit{Classifier and MI Estimator} parts. Through this framework, we explicitly learn disentangled representations and obtain practically good speaker embedding.

The rest of this paper is organized as follows: Section \rom{2} describes the MI estimation and CLUB, and Section \rom{3} presents the proposed framework. Then, the experiments and results are addressed in Sections \rom{4} and \rom{5}, respectively. Finally, we conclude in Section \rom{6}.

\section{Mutual Information Upper Bound Estimation}
Mutual information (MI) is a quantity to measure the amount of dependency between two random variables.
For two continuous random variables $\textbf{x}$ and $\textbf{y}$, MI is defined as follows:
\begin{equation}
\begin{split}
  \mathcal{I}(\textbf{x};\textbf{y}) &= \int p(\textbf{x}, \textbf{y}) \log {p(\textbf{x}, \textbf{y}) \over p(\textbf{x}) p(\textbf{y})} d\textbf{x} d\textbf{y} \\
                           &= \mathbb{E}_{p(\textbf{x}, \textbf{y})} \left[ \log {p(\textbf{x}, \textbf{y}) \over p(\textbf{x}) p(\textbf{y})} \right],
\end{split}
\end{equation}
where $p(\textbf{x}, \textbf{y})$ is the joint distribution, $p(\textbf{x})$ and $p(\textbf{y})$ denote the marginal distributions.

Since our goal is to learn disentangled representations, MI minimization between two random variables is required.
Therefore, we focus on contrastive log-ratio upper bound (CLUB) \cite{20Cheng}, MI upper bound estimator.
For given two random variables $\textbf{x}$ and $\textbf{y}$, CLUB is formulated as follows:
\begin{equation}
\begin{split}
  \mathcal{I}_{\text{CLUB}}(\textbf{x};\textbf{y}) := &\mathbb{E}_{p(\textbf{x}, \textbf{y})} \left[ \log {p(\textbf{y}|\textbf{x})} \right] \\
                                                & - \mathbb{E}_{p(\textbf{x})} \mathbb{E}_{ p(\textbf{y})} \left[ \log {p(\textbf{y}|\textbf{x})} \right].
\end{split}
\end{equation}
As the conditional distribution $p(\textbf{y}|\textbf{x})$ is intractable in our framework, we approximate it using a variational distribution $q_{\phi}(\textbf{y}|\textbf{x})$.
In practice, a variational CLUB (vCLUB) is obtained as follows:
\begin{align}
  &\hat{\mathcal{I}}_{\text{vCLUB}}(\textbf{x};\textbf{y}) = {1 \over N^{2}} \sum_{i=1}^{N} \sum_{j=1}^{N} \Big[ \log {q_{\phi}(\textbf{y}_{i}|\textbf{x}_{i})}  -  \log {q_{\phi}(\textbf{y}_{j}|\textbf{x}_{i})} \Big] \nonumber \\
  & \,\,\,\, = {1 \over N} \sum_{i=1}^{N} \Big[ \log {q_{\phi}(\textbf{y}_{i}|\textbf{x}_{i})} - {1 \over N} \sum_{j=1}^{N} \log {q_{\phi}(\textbf{y}_{j}|\textbf{x}_{i})} \Big],
\end{align}
where $\{(\textbf{x}_{i}, \textbf{y}_{i})\}^{N}_{i=1}$ is $N$ sample pairs drawn from the joint distribution $p(\textbf{x}, \textbf{y})$.
vCLUB is not guaranteed to be the MI upper bound anymore since we approximate $p(\textbf{y}|\textbf{x})$ to $q_{\phi}(\textbf{y}|\textbf{x})$.
However, if the Kullback–Leibler (KL) divergence between conditional and variational distribution is small enough, it can be a reliable MI upper bound estimator. Let $q_{\phi}(\textbf{x},\textbf{y})=q_{\phi}(\textbf{y}|\textbf{x})p(\textbf{x})$ be the variational joint distribution, then KL divergence between $p(\textbf{x}, \textbf{y})$ and $q_{\phi}(\textbf{x},\textbf{y})$ is as follows:
\begin{align}
  &\min_{\phi}\text{KL}(p(\textbf{x}, \textbf{y})||q_{\phi}(\textbf{x},\textbf{y}))  \\
  &= \min_{\phi}\mathbb{E}_{p(\textbf{x}, \textbf{y})} \Big[ \log {p(\textbf{y}|\textbf{x})p(\textbf{x})} - \log {q_{\phi}(\textbf{y}|\textbf{x})p(\textbf{x})} \Big]  \\
  &= \min_{\phi}\mathbb{E}_{p(\textbf{x}, \textbf{y})} \Big[ \log {p(\textbf{y}|\textbf{x})} \Big]  - \mathbb{E}_{p(\textbf{x}, \textbf{y})} \Big[ \log {q_{\phi}(\textbf{y}|\textbf{x})} \Big],
\end{align}
where Equation (6) denotes $\min_{\phi}\text{KL}(p(\textbf{y}|\textbf{x})||q_{\phi}(\textbf{y}|\textbf{x}))$.
Consequently, minimizing $\text{KL}(p(\textbf{y}|\textbf{x})||q_{\phi}(\textbf{y}|\textbf{x}))$ is equivalent to maximizing $\mathbb{E}_{p(\textbf{x}, \textbf{y})} \left[ \log {q_{\phi}(\textbf{y}|\textbf{x})} \right]$ with respect to $\phi$. We train the variational network $q_{\phi}(\textbf{y}|\textbf{x})$ by minimizing the negative log-likelihood loss function as follows:
\begin{align}
  \mathcal{L}_{\text{nll}}(\phi) = - {1 \over N} \sum_{i=1}^{N} \log q_{\phi}(\textbf{y}_{i}|\textbf{x}_{i}).
\end{align}

\begin{figure*}[h]
\begin{minipage}[b]{\linewidth}
  \centering
  \centerline{\includegraphics[width=0.7\linewidth]{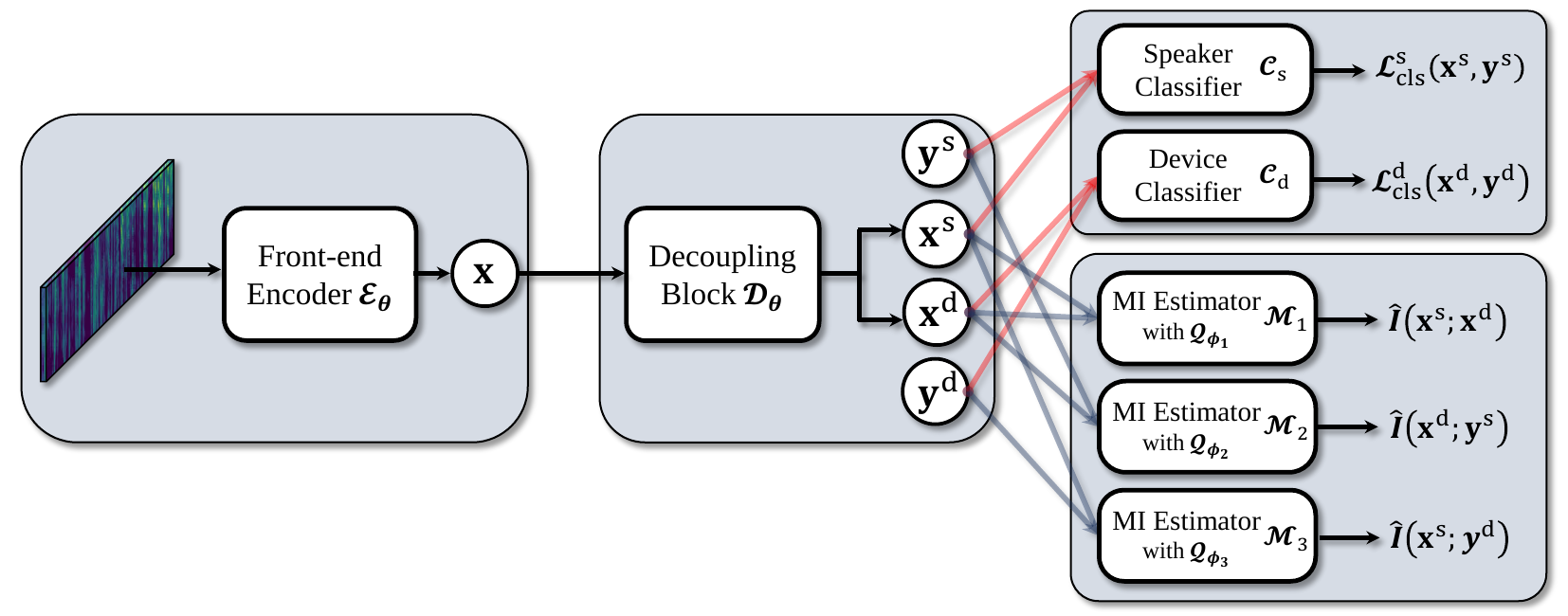}}
\vspace{0cm}
\end{minipage}
\centering\caption{The overall disentangled speaker and device embeddings training framework: Front-end encoder $\mathcal{E}_{\theta}$, decoupling block $\mathcal{D}_{\theta}$, classifiers $\mathcal{C}^{\text{s}}$, $\mathcal{C}^{\text{d}}$ and MI estimators $\mathcal{M}_{1}$, $\mathcal{M}_{1}$, $\mathcal{M}_{3}$ with variational networks $\mathcal{Q}_{\phi_1}$, $\mathcal{Q}_{\phi_2}$, $\mathcal{Q}_{\phi_3}$.}
\label{fig:architecture}
\end{figure*}

\section{Proposed Framework}
In this work, our proposed framework is constructed under the Far-Field Speaker Verification Challenge 2022 (FFSVC 2022) \cite{FFSVC2022_Eval_Plan} scenario to explore more practical cases.
FFSVC 2022 provides a far-field dataset collected by real 155 speakers in complex environments with multiple conditions.
In particular, the datasets of FFSVC 2022 consist of noisy speech samples recorded under far-field conditions and different devices (i.e., tablet, telephone, and microphone array).
In these settings, we learn disentangled speaker and device representations.
As shown in Figure 1, the overall proposed framework is composed of \textit{A. Front-end Encoder Network}, \textit{B. Decoupling Block}, and \textit{C. Classifier and MI Estimator} parts.

\subsection{Front-end Encoder Network}
Given an $F$-dimensional acoustic feature $\textbf{X}\in \mathbb{R}^{F\times T}$ with $T$ frames, the front-end encoder network $\mathcal{E}_{\theta}:\textbf{X} \rightarrow \textbf{x}\in\mathbb{R}^{D}$ extracts an utterance-level initial embedding $\textbf{x}=[x_1,..,x_D]^T$.
To efficiently capture global and local information, we adopt the multi-scale feature aggregation conformer (MFA-Conformer) \cite{zhang2022mfa} backbone and the channel and context-dependent statistic pooling \cite{20Desplanques} for the front-end network.

In case the proposed systems are trained from scratch using a dataset with a limited number of speakers (i.e., FFSVC2022 training set), our disentanglement framework could not work properly (discussed in Section \rom{5}).
To this end, we firstly force the initial embedding to obtain sufficient speaker discrimination ability by pre-training the front-end encoder with a large-scale dataset including many different speakers but no device labels.
Then we fine-tune the whole network with the dataset containing device labels but limited speakers to effectively focus on disentangling the speaker and device factors latent in the initial embedding.

\subsection{Decoupling Block}
To explicitly divide the initial embedding extracted from $\mathcal{E}_{\theta}$ into the latent speaker and device representations, we deploy the decoupling block $\mathcal{D}_{\theta}:\textbf{x} \rightarrow (\textbf{x}^{\text{s}}, \textbf{x}^{\text{d}}) \in \mathbb{R}^{D}$, as shown in Figure 2.
The speaker and device embeddings are obtained via multi-layer perceptron (MLP) modules in $\mathcal{D}_{\theta}$.
MLP module is sequentially comprised of a fully-connected (FC) layer, a batch-normalization (BN) layer, and a rectified linear unit (ReLU) activation function.
Two fixed dimensional embedding vectors, $\textbf{x}^{\text{s}}$ and $\textbf{x}^{\text{d}}$, are learned to represent the input speech's speaker and device characteristics, respectively. For the evaluation, the speaker embeddings $\textbf{x}^{\text{s}}$ are extracted, and the similarities are calculated to perform the verification. 

\subsection{Classifier and MI Estimator}
Analogous to the previous disentanglement approaches \cite{19Zhou, Kang20JFE, 21Zhu, 21Hou, 22Yi}, we follow the multitask learning strategy which includes the classification and MI minimization-based disentanglement tasks.
As shown in Figure 1, the classification task consists of the speaker and device classifiers, $\mathcal{C}_{s}$ and $\mathcal{C}_{d}$, respectively. For the MI minimization-based disentanglement task, there are three MI estimators, $\mathcal{M}_{1}$, $\mathcal{M}_{2}$, and $\mathcal{M}_{3}$.

\vspace{7pt}
\noindent\textbf{Speaker classifier $\mathcal{C}_{\text{s}}$:} To force the speaker embeddings to discriminate their speaker labels, we adopt the combination of the additive angular margin (AAM) softmax \cite{arcface} and the angular prototypical (AP) loss \cite{20Chung}, which has shown the great performance in this field \cite{kwon2021ins, mun2022selective}.
Given the pairs of speaker embeddings and labels $\{(\textbf{x}^{\text{s}}_{i}, \textbf{y}^{\text{s}}_{i})\}^{N}_{i=1}$, the speaker classification loss function is formulated as follows:
\begin{gather}
  \mathcal{L}^{\text{s}}_{\text{AAM}} = - {1 \over N} \sum_{i=1}^{N} \log { e^{s\cos(\vartheta_{\textbf{y}^{\text{s}}_{i}, i}+m)}  \over {e^{s\cos(\vartheta_{\textbf{y}^{\text{s}}_{i}}+m)} + \sum_{j \neq \textbf{y}^{\text{s}}_{i}} e^{s\cos(\vartheta_{j, i})} } }, \\
 \mathcal{L}^{\text{s}}_{\text{AP}} = -{1 \over N} \sum_{i=1}^{N} \log {e^{\text{sim}(\textbf{x}^{\text{s}}_{i,1},\textbf{x}^{\text{s}}_{i,2})}  \over 
  {\sum_{j=1}^{N} e^{\text{sim}(\textbf{x}^{\text{s}}_{i,1},\textbf{x}^{\text{s}}_{j,2})} }},
\end{gather}
\begin{gather}
    \mathcal{L}^{\text{s}}_{\text{cls}} = \mathcal{L}^{\text{s}}_{\text{AAM}} + \mathcal{L}^{\text{s}}_{\text{AP}},
\end{gather}
where $N$ is the batch size, $s$ is a scale factor, $m$ is a margin, $\text{cos}(\vartheta_{j,i})$ is the normalized dot product between the $j$-th class weight of $\mathcal{C}^{\text{s}}$ and $\textbf{x}^{\text{s}}_{i}$, and $\text{sim}(\textbf{x}^{\text{s}}_{i,1}, \textbf{x}^{\text{s}}_{i,2})$ denotes the cosine similarity between two different utterances of $i$-th speaker.

\vspace{8pt}
\noindent\textbf{Device classifier $\mathcal{C}_{\text{d}}$:} As in the speaker classifier, the device embeddings are trained to identify their device labels. The device classification loss is defined as AAM softmax:
\begin{gather}
    \mathcal{L}^{\text{d}}_{\text{cls}} = - {1 \over N} \sum_{i=1}^{N} \log { e^{s\cos(\vartheta_{\textbf{y}^{\text{d}}_{i}, i}+m)}  \over {e^{s\cos(\vartheta_{\textbf{y}^{\text{d}}_{i}}+m)} + \sum_{j \neq \textbf{y}^{\text{d}}_{i}} e^{s\cos(\vartheta_{j, i})} } }.
\end{gather}

\vspace{7pt}
\noindent\textbf{MI estimator $\mathcal{M}_{1}$:} To minimize the MI between speaker and device embeddings, we adopt the mechanism of variational CLUB estimator, which calculates the MI upper bound via the difference of variational distributions between positive and negative sample pairs. The MI upper bound $\hat{\mathcal{I}}(\textbf{x}^{\text{s}};\textbf{x}^{\text{d}})$ between $\textbf{x}^{\text{s}}=[x^{\text{s}}_1,..,x^{\text{s}}_D]^T$ and $\textbf{x}^{\text{d}}=[x^{\text{d}}_1,..,x^{\text{d}}_D]^T$ is estimated as:
\begin{gather}
    \hat{\mathcal{I}}(\textbf{x}^{\text{s}};\textbf{x}^{\text{d}}) = {1 \over N^{2}} \sum_{i=1}^{N} \sum_{j=1}^{N} \Big[ \log { \mathcal{Q}_{\phi_{1}}(\textbf{x}^{\text{s}}_{i}, \textbf{x}^{\text{d}}_{i})  \over  \mathcal{Q}_{\phi_{1}}(\textbf{x}^{\text{s}}_{i}, \textbf{x}^{\text{d}}_{j}) } \Big],
\end{gather}

\begin{figure}[t!]
\begin{minipage}[b]{\linewidth}
  \centering
  \centerline{\includegraphics[width=0.9\linewidth]{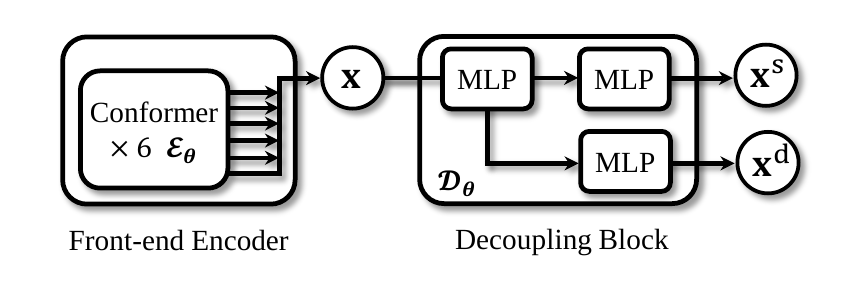}}
\vspace{0.0cm}
\end{minipage}
\centering\caption{Front-end encoder and decoupling block networks.}
\vspace{-0.1cm}
\label{fig:architecture}
\end{figure}

\begin{gather}
    \mathcal{Q}_{\phi_{1}}(\textbf{x}^{\text{s}}, \textbf{x}^{\text{d}}) = \prod_{k=1}^{D} {1\over\sqrt{(2\pi)^D \sigma^{2}_{\phi_{1}}(x^{\text{s}}_k)}}  e^{-{ { (x^{\text{d}}_k - \mu_{\phi_{1}}(x^{\text{s}}_k))^{2} } \over {2 \sigma^{2}_{\phi_{1}}(x^{\text{s}}_k) } } } , 
\end{gather}
where $\mathcal{Q}_{\phi_{1}}$ is the variational network with trainable parameters $\phi_1$ for approximating $p(\textbf{x}^{\text{d}}|\textbf{x}^{\text{s}})$, i.e., representing $q_{\phi_1}(\textbf{x}^{\text{d}}|\textbf{x}^{\text{s}})$. The variational distribution is estimated via the isotropic Gaussian with a diagonal covariance matrix, as shown in Figure 3 (left network).
$\mu_{\phi_1}$ and $\sigma^{2}_{\phi_1} \in \mathbb{R}^{D}$ are obtained via the last two MLP layers of $\mathcal{Q}_{\phi_{1}}$. The parameters of the variational network $\phi_{1}$ are optimized independently with the parameters of the main networks $\theta$ by minimizing the following negative log-likelihood:
\begin{align}
  \mathcal{L}_{\text{nll}_1}(\phi_1) = - {1 \over N} \sum_{i=1}^{N} \log \mathcal{Q}_{\phi_{1}}(\textbf{x}^{\text{s}}_{i}, \textbf{x}^{\text{d}}_{i}).
\end{align}

\vspace{7pt}
\noindent\textbf{MI estimators $\mathcal{M}_{2}$ and $\mathcal{M}_{3}$:} To reduce the interdependence between embeddings and labels, the estimators $\mathcal{M}_{2}$ and $\mathcal{M}_{3}$ estimate the MI upper bounds of $\hat{\mathcal{I}}(\textbf{x}^{\text{d}};\textbf{y}^{\text{s}})$ and $\hat{\mathcal{I}}(\textbf{x}^{\text{s}};\textbf{y}^{\text{d}})$, respectively, through the variational CLUB as follows:
\begin{gather}
    \hat{\mathcal{I}}(\textbf{x}^{\text{d}};\textbf{y}^{\text{s}}) = {1 \over N^{2}} \sum_{i=1}^{N} \sum_{j=1}^{N} \Big[ \log { \mathcal{Q}_{\phi_{2}}(\textbf{x}^{\text{d}}_{i}, \textbf{y}^{\text{s}}_{i})  \over  \mathcal{Q}_{\phi_{2}}(\textbf{x}^{\text{d}}_{i}, \textbf{y}^{\text{s}}_{j}) } \Big], \\
    \log \mathcal{Q}_{\phi_{2}}(\textbf{x}^{\text{d}}, \textbf{y}^{\text{s}}) = - \text{CrossEntropy}(\hat{\textbf{y}}_{\phi_2}(\textbf{x}^{\text{d}}), \textbf{y}^{\text{s}}), \\
    \mathcal{L}_{\text{nll}_2}(\phi_2) = - {1 \over N} \sum_{i=1}^{N} \log \mathcal{Q}_{\phi_{2}}(\textbf{x}^{\text{d}}_{i}, \textbf{y}^{\text{s}}_{i}),
\end{gather}
\begin{gather}    
    \hat{\mathcal{I}}(\textbf{x}^{\text{s}};\textbf{y}^{\text{d}}) = {1 \over N^{2}} \sum_{i=1}^{N} \sum_{j=1}^{N} \Big[ \log { \mathcal{Q}_{\phi_{3}}(\textbf{x}^{\text{s}}_{i}, \textbf{y}^{\text{d}}_{i})  \over  \mathcal{Q}_{\phi_{3}}(\textbf{x}^{\text{s}}_{i}, \textbf{y}^{\text{d}}_{j}) } \Big], \\    
    \log \mathcal{Q}_{\phi_{3}}(\textbf{x}^{\text{s}}, \textbf{y}^{\text{d}}) = - \text{CrossEntropy}(\hat{\textbf{y}}_{\phi_3}(\textbf{x}^{\text{s}}), \textbf{y}^{\text{d}}), \\
    \mathcal{L}_{\text{nll}_3}(\phi_3) = - {1 \over N} \sum_{i=1}^{N} \log \mathcal{Q}_{\phi_{3}}(\textbf{x}^{\text{s}}_{i}, \textbf{y}^{\text{d}}_{i}),
\end{gather}
where $\mathcal{Q}_{\phi_{2}}$ and $\mathcal{Q}_{\phi_{3}}$ are the variational networks with trainable parameters $\phi_2$ and $\phi_3$, respectively, as shown in Figure 3 (right network). $\hat{\textbf{y}}_{\phi}(\textbf{x})$ is the softmax activation output to approximate $p(\textbf{y}|\textbf{x})$. The variational parameters $\phi_2$ and $\phi_3$ are optimized using $\mathcal{L}_{\text{nll}_2}$ and $\mathcal{L}_{\text{nll}_3}$, respectively, in the same way as in the MI estimator $\mathcal{M}_1$.

\subsection{Total Objective Function}
Finally, the main networks (i.e., $\mathcal{E}_{\theta}$, $\mathcal{D}_{\theta}$, $\mathcal{C}^{\text{s}}$, and $\mathcal{C}^{\text{d}}$) are jointly trained with following total objective function:
\begin{equation}
\begin{split} 
    \mathcal{L} = \lambda_{\text{c}_{\text{s}}}\mathcal{L}_{\text{cls}}^{\text{s}} + \lambda_{\text{c}_{\text{d}}} & \mathcal{L}_{\text{cls}}^{\text{d}} + \lambda_{\text{m}_1}\hat{\mathcal{I}}(\textbf{x}^{\text{s}};\textbf{x}^{\text{d}}) \\
    &+\lambda_{\text{m}_2}\hat{\mathcal{I}}(\textbf{x}^{\text{d}};\textbf{y}^{\text{s}}) + \lambda_{\text{m}_3}\hat{\mathcal{I}}(\textbf{x}^{\text{s}};\textbf{y}^{\text{d}}),
\end{split}
\end{equation}
where $\lambda_{\text{c}_{\text{s}}}$, $\lambda_{\text{c}_{\text{d}}}$, $\lambda_{\text{m}_1}$, $\lambda_{\text{m}_2}$, and $\lambda_{\text{m}_3}$ are weighting factors to balance each loss term.
Algorithm 1 summarizes the overall disentangled representation learning framework where $\mathtt{\tiny optimizer}$ is an optimizer, $\eta$ is a learning rate, and $M$ is the number of updates for variatinal networks per epoch. The main and variatinal networks are updated alternately.
\begin{figure}[t!]
\begin{minipage}[b]{\linewidth}
  \vspace{0.15cm}
  \centering
  \centerline{\includegraphics[width=\linewidth]{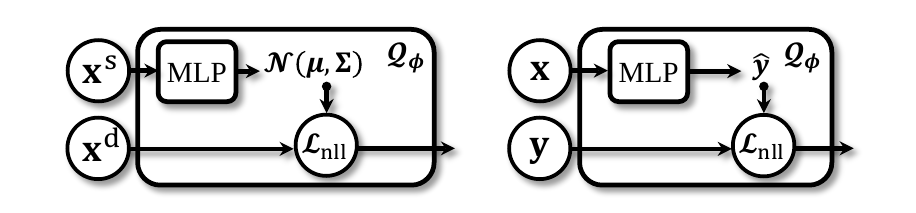}}
\vspace{0cm}
\end{minipage}
\centering\caption{Variational networks. Left network indicates $q_{\phi}(\textbf{x}^{\text{d}}|\textbf{x}^{\text{s}}$) and right network is $q_{\phi}(\textbf{y}|\textbf{x}$).}
\vspace{0.3cm}
\label{fig:architecture}
\end{figure}
\begin{algorithm}[t!]
\caption{Overall Disentangled Speaker and Device Representations Learning Framework.}
\SetAlgoLined
\small{
\KwIn{$ \textbf{X}, \textbf{y}^{\text{s}}, \textbf{y}^{\text{d}}, \, \mathcal{E}_{\theta}, \, \mathcal{D}_{\theta}, \, \mathcal{C}_{\text{s}}, \, \mathcal{C}_{\text{d}}, \, \mathcal{M}_{1}, \, \mathcal{M}_{2}, \, \mathcal{M}_{3}, 
 \break \mathcal{Q}_{\phi_{1}}, \, \mathcal{Q}_{\phi_{2}}, \, \mathcal{Q}_{\phi_{3}}, \, \mathtt{\tiny optimizer}_{\text{m}}, \mathtt{\tiny optimizer}_{\text{v1-3}},
 \break N, \, M, \, \eta_{\text{m}}, \eta_{\text{v1-3}}.$}

\vspace{2mm}
$\mathcal{E}_{\theta}$ is initialized with pre-trained $\theta$. \\

\vspace{2mm}
\For{$k= $ $ 1 $ to $ K$}{
    
    \vspace{1mm}
    $\{(\textbf{X}_{i}, \textbf{y}^{\text{s}}_{i}, \textbf{y}^{\text{d}}_{i})\}_{i=1}^{N} \sim (\textbf{X}, \textbf{y}^{\text{s}}, \textbf{y}^{\text{d}})$ \\

    \vspace{1mm}
    $\{\textbf{x}_{i}\}^{N}_{i=1} \gets \{\mathcal{E}_{\theta}(\textbf{X}_{i})\}^{N}_{i=1}$ \\

    \vspace{1mm}
    $\{(\textbf{x}^{\text{s}}_{i}, \textbf{x}^{\text{d}}_{i})\}^{N}_{i=1} \gets \{\mathcal{D}_{\theta}(\textbf{x}_{i})\}^{N}_{i=1}$ \\

    \vspace{2.5mm}

    \vspace{1mm}
    // \texttt{Variational networks update} \\
    
    \vspace{0.5mm}
    \For{$l= $ $ 1 $ to $ M$}{

        \vspace{1mm}
        \For{$j= $ $ 1 $ to $ N$}{
            \vspace{1mm}
            $\mathcal{L}_{\text{nll}_{1}|j} \gets \mathcal{Q}_{\phi_{1}}(\textbf{x}^{\text{s}}_{j}, \textbf{x}^{\text{d}}_{j})$ \\
    
            \vspace{0.5mm}
            $\mathcal{L}_{\text{nll}_{2}|j} \gets \mathcal{Q}_{\phi_{2}}(\textbf{x}^{\text{d}}_{j}, \textbf{y}^{\text{s}}_{j})$ \\
    
            \vspace{0.5mm}
            $\mathcal{L}_{\text{nll}_{3}|j} \gets \mathcal{Q}_{\phi_{3}}(\textbf{x}^{\text{s}}_{j}, \textbf{y}^{\text{d}}_{j})$ \\
        }
        
        \vspace{0.5mm}
        $\phi_{1} \gets \mathtt{\tiny optimizer}_{\text{v}_1}(\phi_1, \eta_{\text{v}_1}, \nabla_{\phi_1} {1 \over N} \sum_{j=1}^{N} \mathcal{L}_{\text{nll}_{1}|j})$ \\
        
        $\phi_{2} \gets \mathtt{\tiny optimizer}_{\text{v}_2}(\phi_2, \eta_{\text{v}_2}, \nabla_{\phi_2} {1 \over N} \sum_{j=1}^{N} \mathcal{L}_{\text{nll}_{2}|j})$ \\
        
        $\phi_{3} \gets \mathtt{\tiny optimizer}_{\text{v}_3}(\phi_3, \eta_{\text{v}_3}, \nabla_{\phi_3} {1 \over N} \sum_{j=1}^{N} \mathcal{L}_{\text{nll}_{3}|j})$ \\
    }

    \vspace{2mm}
    // \texttt{Main networks update} \\
    
    \vspace{0.5mm}
    \For{$i= $ $ 1 $ to $ N$}{
        \vspace{1mm}
        $\mathcal{L}_{\text{cls}|i}^{\text{s}} \gets \mathcal{C}_{\text{s}}(\textbf{x}^{\text{s}}_{i}, \textbf{y}^{\text{s}}_{i})$ \\
    
        \vspace{0.5mm}
        $\mathcal{L}_{\text{cls}|i}^{\text{d}} \gets \mathcal{C}_{\text{d}}(\textbf{x}^{\text{d}}_{i}, \textbf{y}^{\text{d}}_{i})$ \\
    
        \vspace{1mm}
        $\hat{\mathcal{I}}(\textbf{x}^{\text{s}}_{i}; \textbf{x}^{\text{d}}_{i}) \gets \mathcal{M}_{1}(\textbf{x}^{\text{s}}_{i}, \textbf{x}^{\text{d}}_{i}, \mathcal{Q}_{\phi_{1}})$ \\
    
        \vspace{1mm}
        $\hat{\mathcal{I}}(\textbf{x}^{\text{d}}_{i}; \textbf{y}^{\text{s}}_{i}) \gets \mathcal{M}_{2}(\textbf{x}^{\text{d}}_{i}, \textbf{y}^{\text{s}}_{i}, \mathcal{Q}_{\phi_{2}})$ \\
    
        \vspace{1mm}
        $\hat{\mathcal{I}}(\textbf{x}^{\text{s}}_{i}; \textbf{y}^{\text{d}}_{i}) \gets \mathcal{M}_{3}(\textbf{x}^{\text{s}}_{i}, \textbf{y}^{\text{d}}_{i}, \mathcal{Q}_{\phi_{3}})$ \\
    
        \vspace{1mm}
        $\mathcal{L}_{i} \gets \mathcal{L}_{\text{cls}|i}^{\text{s}} + \
                             \mathcal{L}_{\text{cls}|i}^{\text{d}} + \
                             \hat{\mathcal{I}}(\textbf{x}^{\text{s}}_{i}; \textbf{x}^{\text{d}}_{i}) + \
                             \hat{\mathcal{I}}(\textbf{x}^{\text{d}}_{i}; \textbf{y}^{\text{s}}_{i}) + \
                             \hat{\mathcal{I}}(\textbf{x}^{\text{s}}_{i}; \textbf{y}^{\text{d}}_{i})$ \\
    }
    
    \vspace{1mm}
    $\theta \gets \mathtt{\tiny optimizer}_{\text{m}}(\theta, \eta_{\text{m}}, \nabla_{\theta} {1 \over N} \sum^{N}_{i=1} \mathcal{L}_{i})$
    }   
}
\end{algorithm}
\begin{table*}[t!]
\centering
\caption{Speaker verification performances on the FFSVC 2022 development trial protocol. $\dagger$: Our re-implementation.}
\label{tab1:table}
\renewcommand{\tabcolsep}{2.5mm}
\small{
\begin{tabular}{ccccc}
\toprule[.1em]
      \multicolumn{1}{c}{\multirow{2}{*}{\begin{tabular}[c]{@{}c@{}}\textbf{Pre-training Dataset} \\ \textbf{(Front-end Encoder)} \end{tabular}}} & 
      \multicolumn{1}{c}{\multirow{2}{*}{\begin{tabular}[c]{@{}c@{}}\textbf{Fine-tuning Dataset} \\ \textbf{(whole network)} \end{tabular}}} & 
      \multicolumn{1}{c}{\multirow{2}{*}{\begin{tabular}[c]{@{}c@{}}\textbf{Objective Function}\end{tabular}}} & 
      \multicolumn{2}{c}{\textbf{Development Set}}    \\
      \cmidrule(lr){4-5}
       \multicolumn{1}{c}{} &
       \multicolumn{1}{c}{} &
       \multicolumn{1}{c}{} &
       \multicolumn{1}{c}{\textbf{EER(\%)}} &
       \multicolumn{1}{c}{\textbf{MinDCF}} \\
       
\midrule[.08em]

    VoxCeleb      & $\times$     & \multirow{1}{*}{\textit{Only using initial embedding from pre-trained $\mathcal{E}_{\theta}$}} & 12.09 & 0.722 \\
    \arrayrulecolor{black!40}\midrule[.04em]

    $\times$      & FFSVC2022   & \multirow{2}{*}{JFE$^\dagger$ \cite{Kang20JFE}}          & 11.98  & 0.688 \\ 
    VoxCeleb      & FFSVC2022   &                                                          & 7.02   & 0.460 \\ 
    \arrayrulecolor{black!100}
    \specialrule{0.8pt}{2.0pt}{2.0pt}
    \specialrule{0.8pt}{0.0pt}{3.0pt}

    $\times$      & FFSVC2022   & \multirow{2}{*}{$\mathcal{L}_{\text{cls}}^{\text{s}}$} & 11.83  & 0.668 \\ 
    VoxCeleb      & FFSVC2022   &                                                          & 7.08   & 0.468 \\ 
    \arrayrulecolor{black!40}\midrule[.04em]

    $\times$      & FFSVC2022   & \multirow{2}{*}{$\mathcal{L}_{\text{cls}}^{\text{s}} + \mathcal{L}_{\text{cls}}^{\text{d}}$} & 12.20 & 0.690 \\
    VoxCeleb      & FFSVC2022   &                                                                                                  & 7.15  & 0.473 \\
    \arrayrulecolor{black!40}\midrule[.04em]

    $\times$      & FFSVC2022   & \multirow{2}{*}{$\mathcal{L}_{\text{cls}}^{\text{s}} 
                                                  +\mathcal{L}_{\text{cls}}^{\text{d}} 
                                                  +\hat{\mathcal{I}}(\textbf{x}^{\text{s}};\textbf{x}^{\text{d}})$}               & 12.06 & 0.718 \\
    VoxCeleb      & FFSVC2022   &                                                                                                  & 7.03  & 0.467 \\ 
    \arrayrulecolor{black!40}\midrule[.04em]

    $\times$      & FFSVC2022   & \multirow{2}{*}{$\mathcal{L}_{\text{cls}}^{\text{s}} 
                                                  +\mathcal{L}_{\text{cls}}^{\text{d}} 
                                                  +\hat{\mathcal{I}}(\textbf{x}^{\text{d}};\textbf{y}^{\text{s}})
                                                  +\hat{\mathcal{I}}(\textbf{x}^{\text{s}};\textbf{y}^{\text{d}})$}               & 12.00 & 0.703 \\
    VoxCeleb      & FFSVC2022   &                                                                                                  & 6.99  & 0.461 \\ 
    \arrayrulecolor{black!40}\midrule[.04em]

    $\times$      & FFSVC2022   & \multirow{2}{*}{$\mathcal{L}_{\text{cls}}^{\text{s}} 
                                                  +\mathcal{L}_{\text{cls}}^{\text{d}} 
                                                  +\hat{\mathcal{I}}(\textbf{x}^{\text{s}};\textbf{x}^{\text{d}})
                                                  +\hat{\mathcal{I}}(\textbf{x}^{\text{d}};\textbf{y}^{\text{s}})
                                                  +\hat{\mathcal{I}}(\textbf{x}^{\text{s}};\textbf{y}^{\text{d}})$}               & 11.95 & 0.684 \\
    \textbf{VoxCeleb} & \textbf{FFSVC2022}&                                                                                        & \textbf{6.95} & \textbf{0.450} \\ 
    
\arrayrulecolor{black}\bottomrule[.1em]
\end{tabular}
} \end{table*}

\section{Experiments}
\subsection{Datasets}
To pre-train the front-end encoder network $\mathcal{E}_{\theta}$, we employ the development set of VoxCeleb1 and VoxCeleb2 datasets \cite{17Nagrani, 18Chung, 20Nagrani}, which consist of 1,092,009 and 148,642 utterances from 5,994 and 1,211 speakers, respectively.
VoxCeleb dataset is one of the most popular corpora for large-scale text-independent speaker verification.
The speech samples were extracted from YouTube video clips and degraded with real-world noises, including background chatter, laughter, overlapping speech, room acoustics, etc.
The front-end encoder network was trained in a fully supervised learning manner with the speaker classifier.

When fine-tuning the whole network with pre-trained $\mathcal{E}_{\theta}$, we use the FFSVC2022 training dataset which is the composition of the training, development, and supplementary sets of the FFSVC 2020 challenge \cite{FFSVC2020}.
FFSVC2022 training dataset totally contains 2,548,351 utterances from 155 speakers where we only utilize samples longer than 1 second (i.e., 2,542,392 utterances).
FFSVC2022 dataset was collected from four recording devices (i.e., iPhone, Android phone, iPad, and normal/circular microphone array) in six different locations (i.e., 0m, 25cm, 1m, 1.5m, 3m, and 5m).
For our disentangled representation learning framework, we fine-tuned the whole network using the utterances with corresponding speaker and device labels.

\begin{figure*}[t!]
     \centering
     \begin{subfigure}[b]{0.75\textwidth}
         \includegraphics[width=0.49\textwidth]{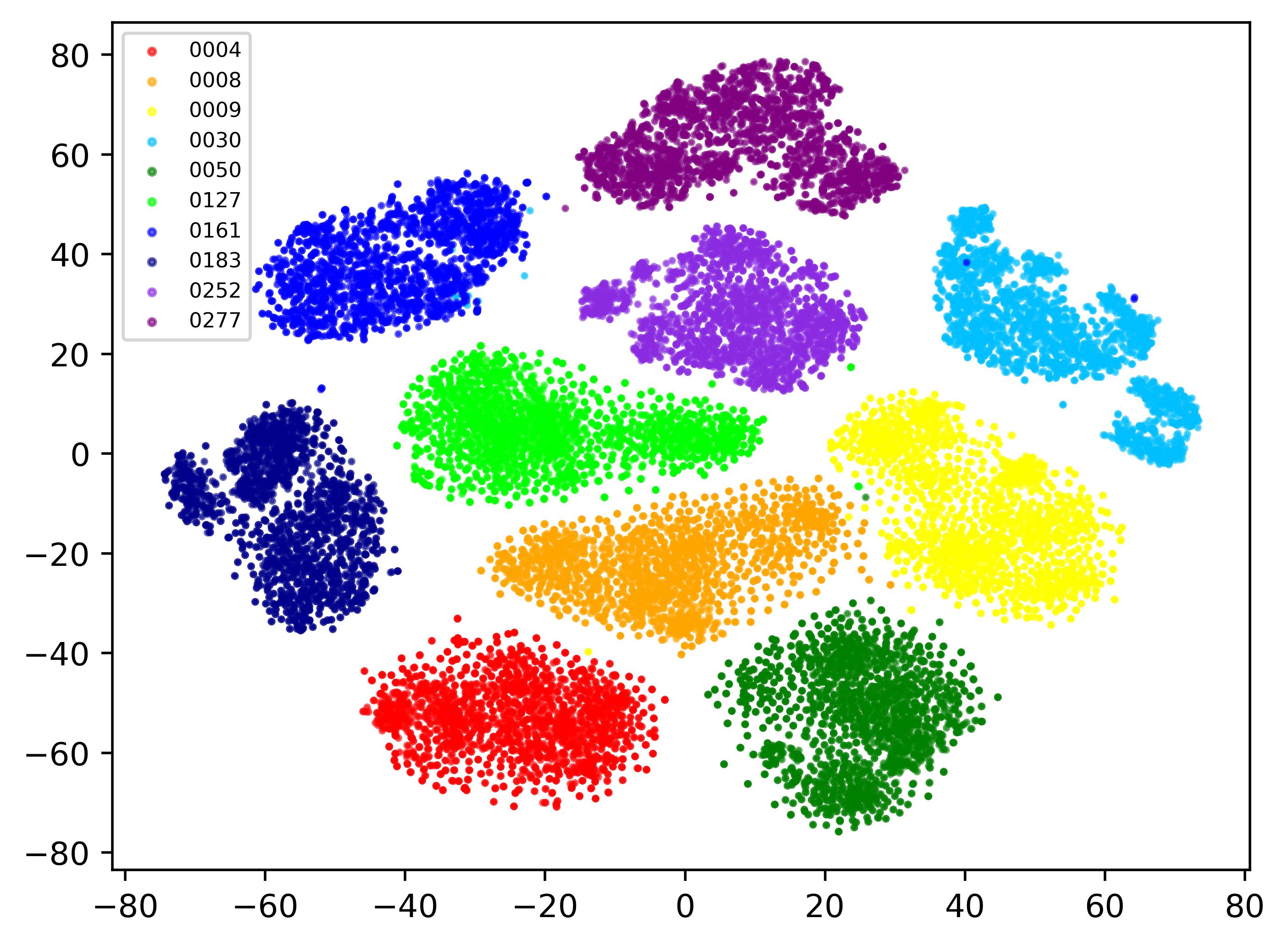} 
         \hfill
         \includegraphics[width=0.49\textwidth]{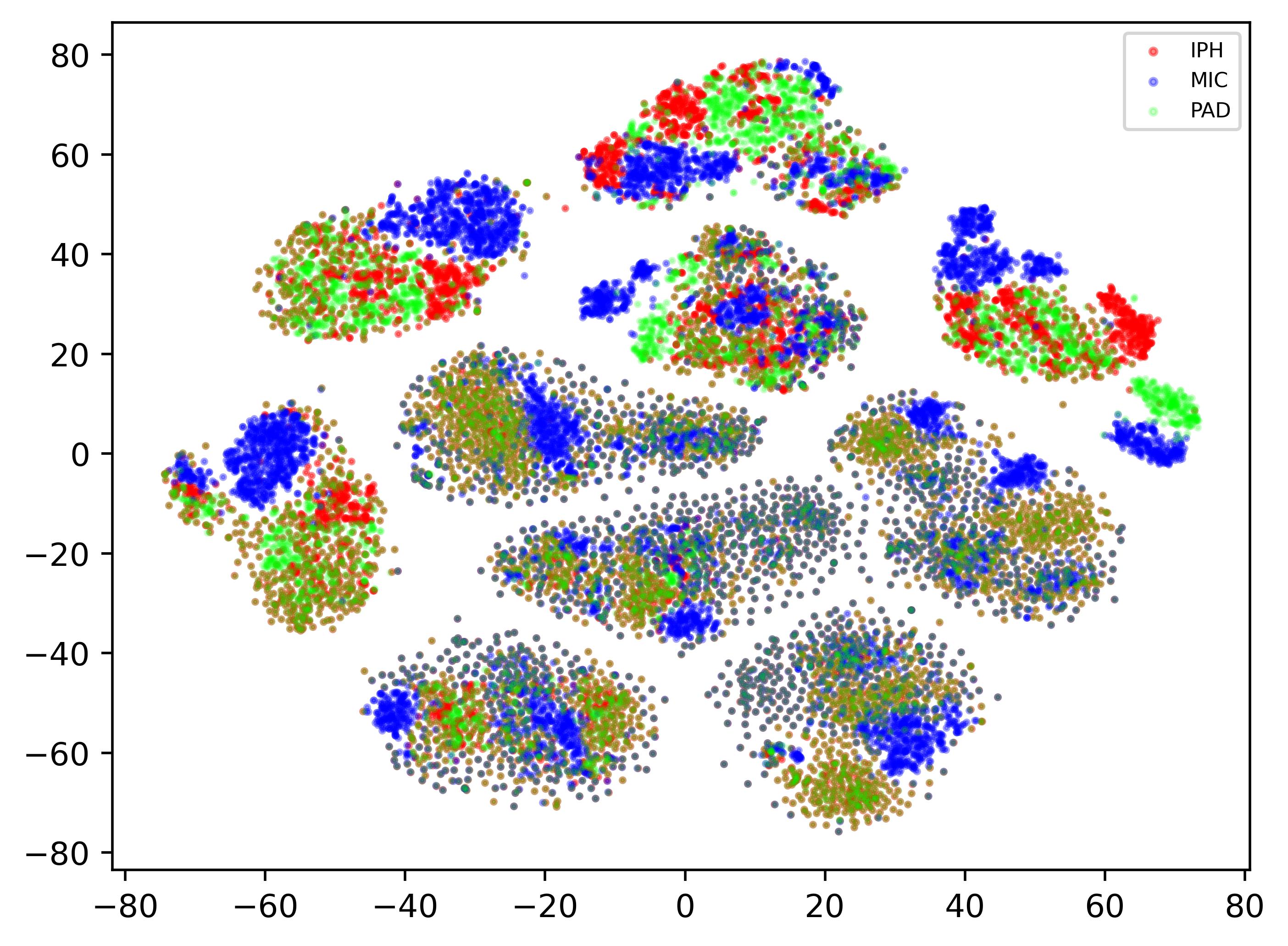} 
         \vspace{-0.15cm}
         \caption{Only speaker classification objective function: $\mathcal{L}_{\text{cls}}^{\text{s}}=\mathcal{L}^{\text{s}}_{\text{AAM}} + \mathcal{L}^{\text{s}}_{\text{AP}}$.}
     \end{subfigure}
     \vspace{0.4cm}
     
     \begin{subfigure}[b]{0.75\textwidth}
         \includegraphics[width=0.49\textwidth]{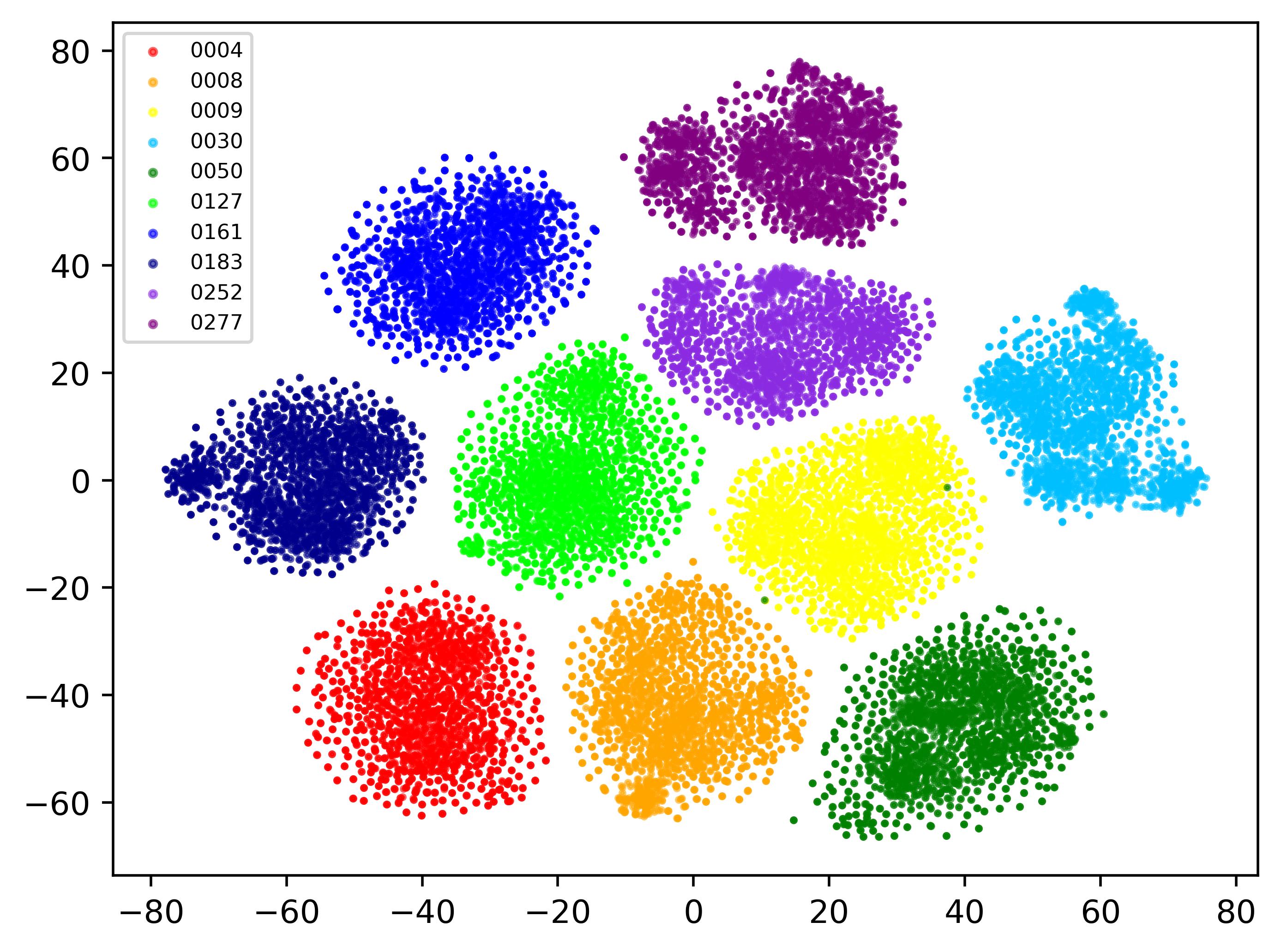}
         \hfill
         \includegraphics[width=0.49\textwidth]{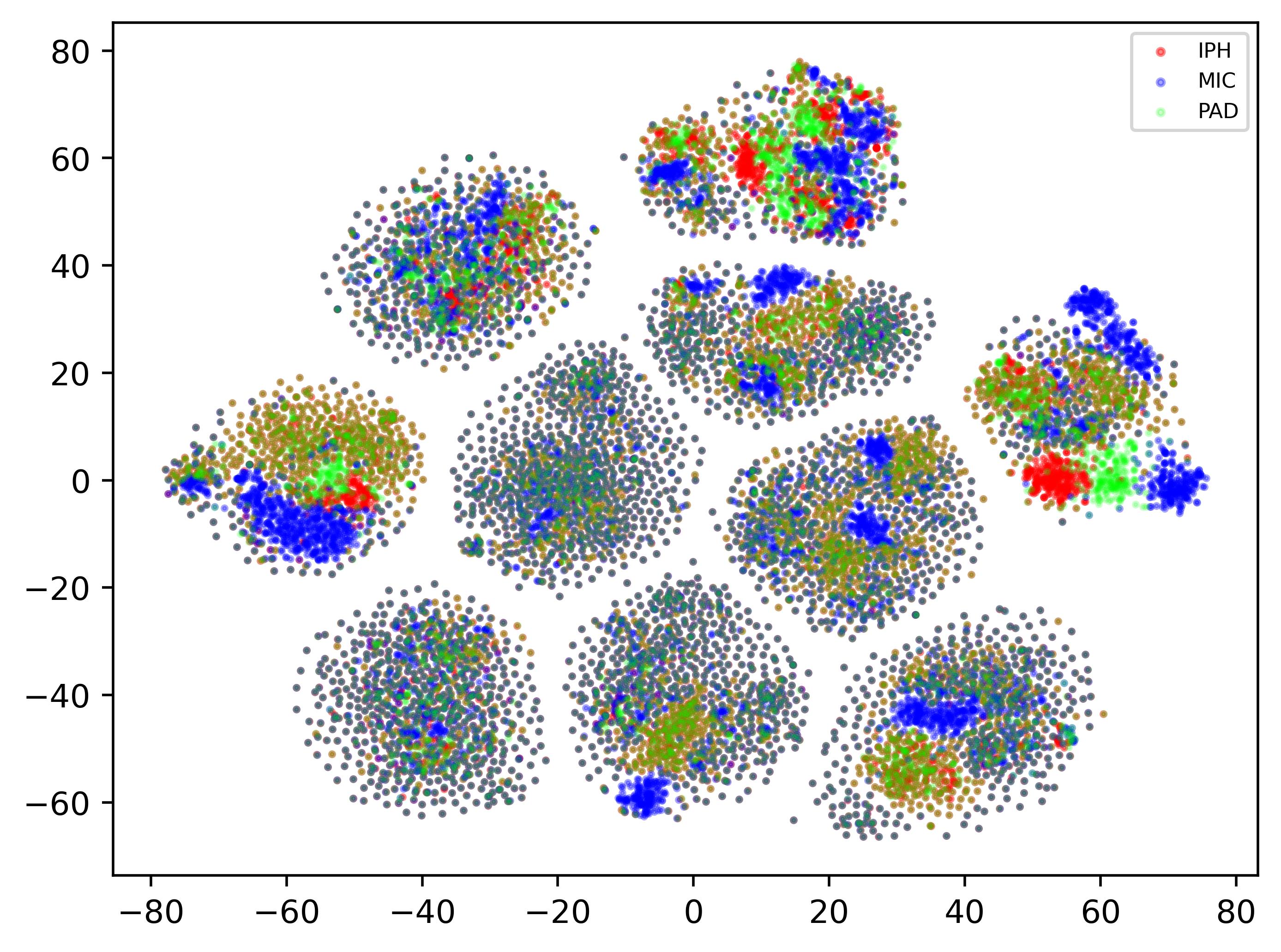}
         \vspace{-0.15cm}
         \caption{JFE objective function \cite{Kang20JFE}: $\mathcal{L}_{\text{s-s,CE}} + \mathcal{L}_{\text{d-d,CE}} - \mathcal{L}_{\text{s-d,E}} - \mathcal{L}_{\text{d-s,E}} - \mathcal{L}_{\text{nMAPC}}$.}
     \end{subfigure}
     \vspace{0.4cm}
    
     \begin{subfigure}[b]{0.75\textwidth}
         \includegraphics[width=0.49\textwidth]{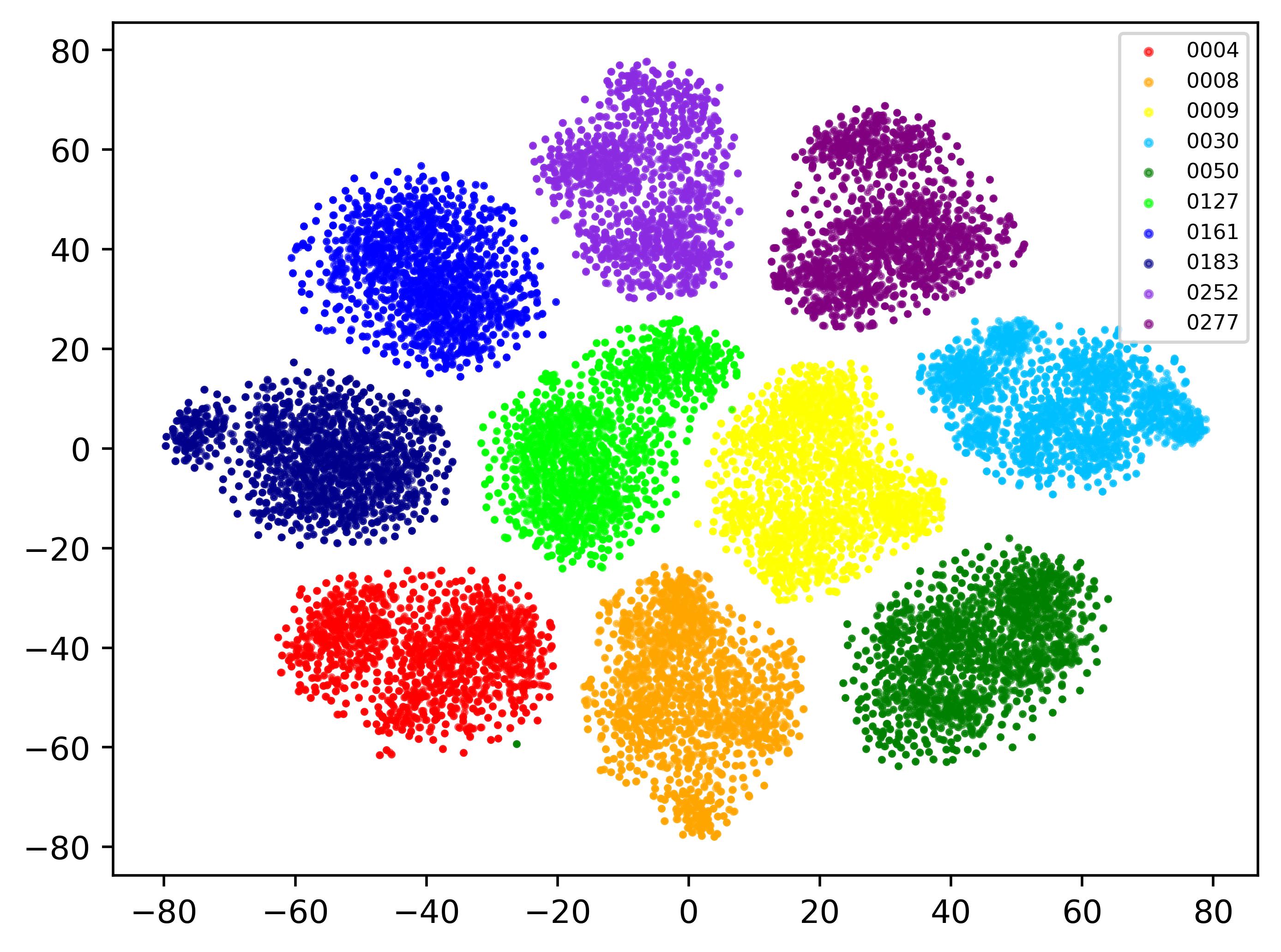} 
         \hfill
         \includegraphics[width=0.49\textwidth]{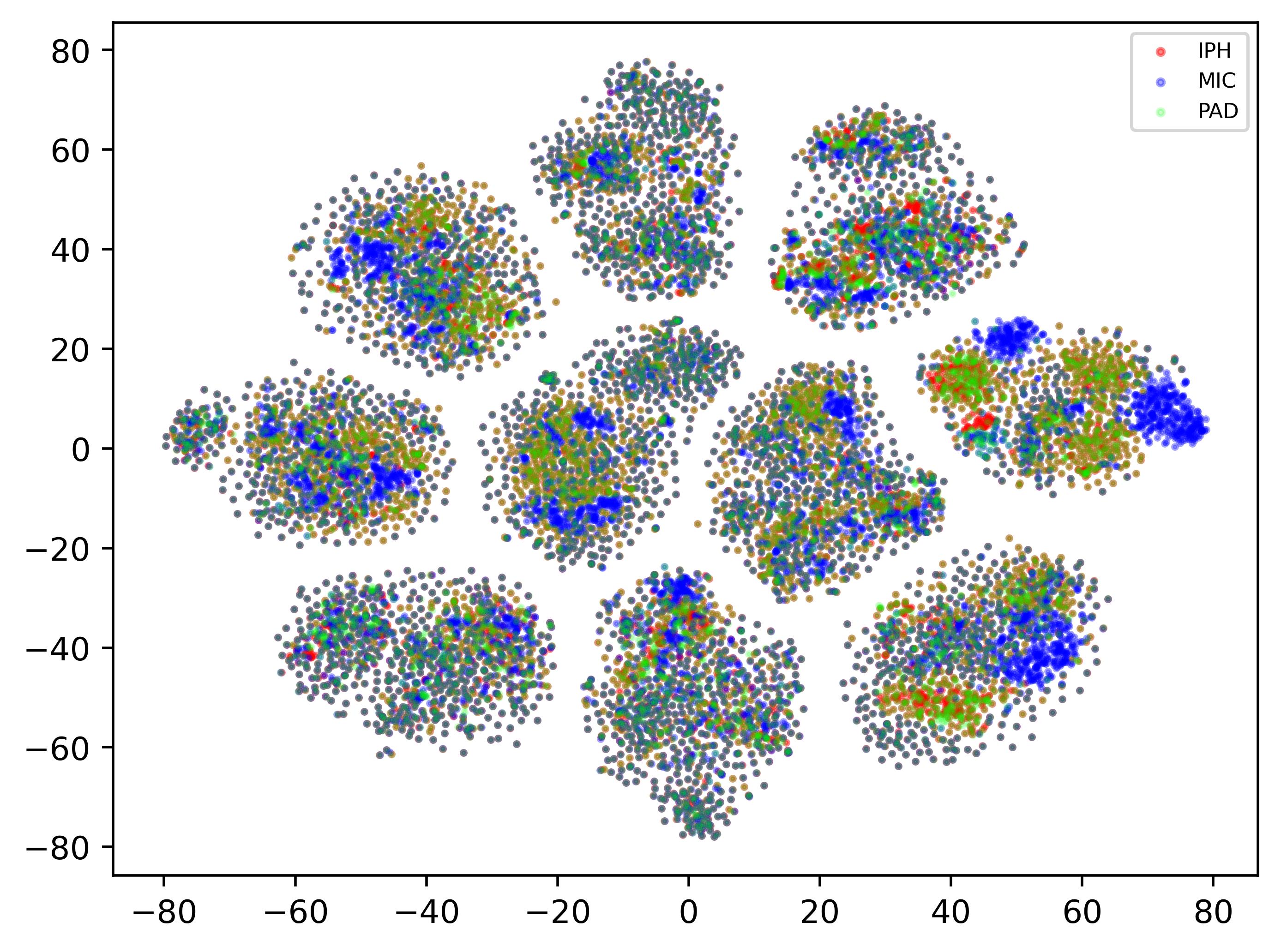} 
         \vspace{-0.15cm}
         \caption{Proposed objective function: $\mathcal{L}_{\text{cls}}^{\text{s}}+\mathcal{L}_{\text{cls}}^{\text{d}}+\hat{\mathcal{I}}(\textbf{x}^{\text{s}};\textbf{x}^{\text{d}})+\hat{\mathcal{I}}(\textbf{x}^{\text{d}};\textbf{y}^{\text{s}})+\hat{\mathcal{I}}(\textbf{x}^{\text{s}};\textbf{y}^{\text{d}}$).}
     \end{subfigure}
     \vspace{0.3cm}
        \caption{t-SNE visualization of speaker embedding space results. The input utterances are randomly sampled from FFSVC2022 dataset. The sampled utterances consist of ten speakers (\texttt{\footnotesize 0004,0008,0009,0030,0050,0127,0161,0183,0252,0277}) under three recording devices (iPhone-\texttt{IPH}, microphone array-\texttt{MIC}, and iPad-\texttt{IPD}). Speaker embeddings $\textbf{x}^{\text{s}}$ with speaker labels $\textbf{y}^{\text{s}}$ (left) and device labels $\textbf{y}^{\text{d}}$ (right) are learned using (a) only speaker classification loss, (b) JFE \cite{Kang20JFE}, and (c) proposed objective function.}
        \label{fig:three graphs}
\end{figure*}

\subsection{Evaluation Protocol and Metrics}
To evaluate the system performance, we adopt development trial protocol provided by FFSVC2022 challenge, which was utilized to tune hyper-parameters and validate the model performance during the previous competition period \cite{FFSVC2022_Eval_Plan}. Since the FFSVC2022 development trial protocol contains speech samples collected by real speakers in multiple environments, we can evaluate the system performance in realistic scenarios with multiple conditions.
We report two performance metrics: the equal error rate (EER) and the minimum detection cost function (MinDCF). The EER is the error when the false alarm rate (FAR) and the false reject rate (FRR) are the same, and the MinDCF is defined as the minimum value of the weighted sum of the FAR and FRR.
The parameters of MinDCF were set as $C_{\text{miss}}=1$, $C_{\text{fa}}=1$, and $P_{\text{target}}=0.05$.

\subsection{Model Architectures}
For the front-end encoder network, we adopt the MFA-Conformer \cite{zhang2022mfa} architecture, which is the multi-scale feature-aggregated encoder for extracting speaker embedding based on the convolution-augmented transformer.
We use six conformer layers which consist of the multi-headed self-attention module (MHSA), the convolution module (CM), the feed-forward module (FFM), and the sub-sampling layer (SSL).
For the MHSA, the encoder dimension, the number of attention heads, the dropout rate, and the kernel size are set to 256, 4, 0.1, and 15, respectively.
For the CM, the kernel size is set to 15.
For the FFM, FC layers with the dimension of 2,048 are used.
For the SSL, a convolution layer with a sub-sampling rate of 2 is employed.
We aggregate the frame-level output features to the 192-dimensional initial embedding $\textbf{x}$ via the channel and context-dependent statistic pooling \cite{20Desplanques}.
In the decoupling block, there are three MLP layers, as shown in Figure 2, where each MLP layer consists of FC-ReLU-BN sequentially.
From the outputs of the last two MLP layers, the 192-dimensional speaker and device embeddings are obtained.
The dimension of the hidden and last FC layers for the variational network $\mathcal{Q}_{\phi_1}$ is set to 1,024 and 192, respectively.

\subsection{Baseline: Joint Factor Embedding (JFE)}
To compare with the existing disentanglement method, we adopt joint factor embedding (JFE) \cite{Kang20JFE}.
JFE framework simultaneously learns speaker and nuisance (device) embeddings where the cross-entropy on their main task ($\mathcal{L}_{\text{s-s,CE}}$ and $\mathcal{L}_{\text{d-d,CE}}$) is minimized while the entropy on their opposite task ($\mathcal{L}_{\text{s-d,E}}$ and $\mathcal{L}_{\text{d-s,E}}$) is maximized.
Also, the negative MAPC between two embeddings ($\mathcal{L}_{\text{nMAPC}}$) is jointly minimized.
For our experimental setting, the speaker and device embeddings are optimized using the following JFE objective function:
\begin{equation}
\begin{split} 
    \mathcal{L}_{\text{JFE}} = \lambda_{\text{s-s}} & \mathcal{L}_{\text{s-s,CE}} + \lambda_{\text{d-d}}\mathcal{L}_{\text{d-d,CE}} \\
    &- \lambda_{\text{s-d}}\mathcal{L}_{\text{s-d,E}} - \lambda_{\text{d-s}}\mathcal{L}_{\text{d-s,E}} - \lambda_{\text{nMAPC}}\mathcal{L}_{\text{nMAPC}}.
\end{split}
\end{equation}

\subsection{Implementation Details}
We made use of the PyTorch library and conducted experiments using $2$ NVIDIA GeForce RTX 3090 GPUs in parallel\footnote{All implementations are developed based on \texttt {\url{https://github.com/clovaai/voxceleb_trainer}.}}.
During both pre-training and fine-tuning phases, we randomly cropped an input utterance to 200-frames segment and then applied MUSAN noises \cite{15Snyder} or the simulated room impulse responses (RIRs) \cite{17Ko} for data augmentation.
If input utterance is shorter than 200 frames, we duplicated and randomly selected 200-frames segment.
Acoustic features are 80-dimensional log mel-filterbanks with a hamming window length of 25ms and hop-size of 10ms with 512-size FFT bins. 
Mean and variance normalization is applied to the log mel-filterbanks.
The AAM-softmax loss function~\cite{arcface} employs a margin of 0.2 and a scale of 30.
The AP loss function~\cite{20Chung} uses the prototype with one utterance.
We adopted a batch size of $200$ and an Adam optimizer with a weight decay of 2e-5.
For the pre-training phase, we scheduled the learning rate via the cosine annealing with warm-up restart (SGDR) \cite{loshchilov2016sgdr} with a cycle size of 25 epochs, the maximum learning rate of 1e-3 and the decreasing rate of 0.8 for two cycles.
In the fine-tuning phase, we set the hyper-parameters of SGDR scheduler to a cycle size of 4 epochs, the maximum learning rate of 1e-5, and the minimum learning rate of 1e-8 for one cycle.
The weighting factors for total objective function are set to $\lambda_{\text{c}_{\text{s}}}=5$, $\lambda_{\text{c}_{\text{d}}}=10$, $\lambda_{\text{m}_1}=0.5$, $\lambda_{\text{m}_2}=0.1$, and $\lambda_{\text{m}_3}=0.1$.
The weighting factors for JFE objective function are set to $\lambda_{\text{s-s}}=\lambda_{\text{d-d}}=1$, $\lambda_{\text{s-d}}=\lambda_{\text{d-s}}=0.00001$, and $\lambda_{\text{nMAPC}}=0.0001$.

\section{Results}
\subsection{Speaker Verification Performance}
Table 1 shows the speaker verification performances on the FFSVC 2022 development set.
We report the experimental results of seven systems to compare the verification performance of proposed methods with the baseline and analyze the effect of each objective function term in the proposed framework, i.e., $\mathcal{L}_{\text{cls}}^{\text{s}}$, $\mathcal{L}_{\text{cls}}^{\text{d}}$, $\hat{\mathcal{I}}(\textbf{x}^{\text{s}};\textbf{x}^{\text{d}})$, $\hat{\mathcal{I}}(\textbf{x}^{\text{d}};\textbf{y}^{\text{s}})$, and $\hat{\mathcal{I}}(\textbf{x}^{\text{s}};\textbf{y}^{\text{d}})$.

In Table 1, the first row shows the result using the only initial embedding $\textbf{x}$ from the pre-trained front-end encoder $\mathcal{E}_{\theta}$ without fine-tuning.
The second row in Table 1 indicates the performance of the JFE baseline described in Section \rom{4}.D.
The systems from the third to seventh rows in Table 1 show the results using the speaker embeddings $\textbf{x}^{\text{s}}$ fine-tuned with (3$^{\text{rd}}$ row) the speaker classification loss $(\mathcal{L}_{\text{cls}}^{\text{s}})$, 
(4$^{\text{th}}$ row) the multi-task learning of speaker and device classification losses $(\mathcal{L}_{\text{cls}}^{\text{s}}+\mathcal{L}_{\text{cls}}^{\text{d}})$, 
(5$^{\text{th}}$ row) the multi-task learning including the estimated MI between $\textbf{x}^{\text{s}}$ and $\textbf{x}^{\text{d}}$ loss $(\mathcal{L}_{\text{cls}}^{\text{s}}+\mathcal{L}_{\text{cls}}^{\text{d}}+\hat{\mathcal{I}}(\textbf{x}^{\text{s}};\textbf{x}^{\text{d}}))$, 
(6$^{\text{th}}$ row) the multi-task learning including the estimated MIs between the embeddings and labels loss $(\mathcal{L}_{\text{cls}}^{\text{s}}+\mathcal{L}_{\text{cls}}^{\text{d}}+\hat{\mathcal{I}}(\textbf{x}^{\text{d}};\textbf{y}^{\text{s}})+\hat{\mathcal{I}}(\textbf{x}^{\text{s}};\textbf{y}^{\text{d}}))$, and 
(7$^{\text{th}}$ row) the total objective loss $(\mathcal{L}_{\text{cls}}^{\text{s}}+\mathcal{L}_{\text{cls}}^{\text{d}}+\hat{\mathcal{I}}(\textbf{x}^{\text{s}};\textbf{x}^{\text{d}})+\hat{\mathcal{I}}(\textbf{x}^{\text{d}};\textbf{y}^{\text{s}})+\hat{\mathcal{I}}(\textbf{x}^{\text{s}};\textbf{y}^{\text{d}}))$.
Also, for each system, we report the results of fine-tuning with a randomly initialized front-end encoder $\mathcal{E}_{\theta}$ from scratch.

As shown in Table 1, where upper values in each row indicate the performances without pre-training, applying regularization terms, i.e., multi-task learning $(\mathcal{L}_{\text{cls}}^{\text{d}})$ and CLUB estimators $(\hat{\mathcal{I}}(\textbf{x}^{\text{s}};\textbf{x}^{\text{d}})
,\, \hat{\mathcal{I}}(\textbf{x}^{\text{d}};\textbf{y}^{\text{s}})
,\, \hat{\mathcal{I}}(\textbf{x}^{\text{s}};\textbf{y}^{\text{d}}))$, did not show significant improvement in the speaker verification performance but rather even degrades the system.
However, utilizing the front-end encoder $\mathcal{E}_{\theta}$ pre-trained using a large-scale dataset without the device labels significantly improved the system performance.
This shows that the proposed framework can work effectively when the speaker and device factors, $\textbf{x}^{\text{s}}$ and $\textbf{x}^{\text{d}}$, latent in the shared embedding $\textbf{x}$ are separated after securing sufficient speaker discrimination ability.
Comparing the 3$^\text{rd}$ and 4$^\text{th}$ rows of Table 1 in the cases using the pre-trained $\mathcal{E}_{\theta}$, we observed that multi-task learning does not help improve the verification performance.
However, jointly employing the MI regularization terms led to a consistent performance improvement, as shown in the 5$^\text{th}$, 6$^\text{th}$ and 7$^\text{th}$ rows.
Finally, we obtained the best performing result using the final objective function, achieving EER of 6.95\% and MinDCF of 0.450 on the FFSVC2022 development trial protocol, respectively.
These results outperform those of the JFE baseline system of the 2$^\text{nd}$ row in Table 1.

\subsection{Visualization of Speaker Embedding Space}
We also investigate the effect of our proposed framework in embedding space by visualizing the speaker representations learned using the three different training strategies, i.e., (a) only speaker classification loss, (b) JFE objective function \cite{Kang20JFE}, and (c) the proposed objective function.
Figure 4 (a), (b), and (c) show the t-SNE plots of speaker embeddings of ten speakers and three devices.
Embedding points are colored by speaker labels in the left parts of Figure 4 while colored by device labels in the right parts.

As shown in the left parts of Figure 4 (a), (b), and (c), the speaker embeddings are well separated between different speakers. However, from the view of the device label in the right parts of Figure 4, the embedding points of different devices are highly overlapped, making it difficult to identify their own color (red, blue, and green).
In particular, it is observed that the embedding points in the right part of Figure 4 (c) are more evenly dispersed over different devices compared to those in the right parts of Figure 4 (a) and (b).
This shows that the speaker embedding extracted from the proposed framework is well-discriminated in the main task while indistinguishable in the sub-task. 
Furthermore, the speaker embedding learned via our proposed framework demonstrates a more disentangled visualization result than the speaker embeddings obtained from other training strategies, i.e., only speaker classification loss and JFE objective function.

\section{Conclusion}
In this paper, we propose a novel framework for disentangling speaker representation from speaker-irrelevant factors in a direct manner. The proposed framework can explicitly reduce the mutual information by minimizing the estimation of its upper bound. Through mutual information minimization, the interdependence of decoupled speaker and device embedding is removed. Experimental results demonstrate that our approach can improve the speaker verification performance by taking advantage of the pre-trained front-end encoder. Also, visualization of speaker embedding space shows that device-dependent factor in speaker embedding is dispersed, from which we can assert that their inter-dependency is lost.

\section*{Acknowledgment}
This work was supported by Institute of Information \& communications Technology Planning \& Evaluation (IITP) grant funded by the Korea government (MSIT) (No.2021-0-00456, Development of Ultra-high Speech Quality  Technology for Remote Multi-speaker Conference System).

\end{document}